\documentclass[reprint,twocolumn,floats,aip,jcp,english,preprintnumbers,amsmath,amssymb]{revtex4-1}
\usepackage{graphicx,amsmath}
\usepackage{dcolumn}

\begin{document}
\author{Hagai Eshet}
\affiliation{$\dagger$School of Chemistry, The Sackler Faculty of Exact Sciences,
  Tel Aviv University, Tel Aviv 69978, Israel}
\author{Michael Gr\"unwald}
\affiliation{$\ddagger$Computational Physics, University of Vienna, Sensengasse
  8, 1090 Vienna, Austria}
\author{Eran Rabani}
\affiliation{$\dagger$School of Chemistry, The Sackler Faculty of Exact Sciences,
  Tel Aviv University, Tel Aviv 69978, Israel}
\email{eran.rabani@gmail.com}

\title[]{The Electronic Structure of CdSe/CdS Core/Shell Seeded
  Nanorods: Type-I or Quasi-Type-II?}

\begin{abstract}
The electronic structure of CdSe/CdS core/shell seeded nanorods of
experimentally relevant size is studied using a combination of
molecular dynamics and semiempirical pseudopotential techniques, with
the aim to address the transition from type-I to a quasi-type-II band
alignment. The hole is found to be localized in the core region
regardless of its size. The overlap of the electron density with the
core region depends markedly on the size of the CdSe core: For small
cores, we observe little overlap, consistent with type-II
behavior. For large cores, significant core-overlap of a number of
excitonic states can lead to type-I behavior. When electron-hole
interactions are taken into account, the core-overlap is further
increased. Our calculations indicate that the observed transition from
type-II to type-I is largely due to simple volume effects, and not to
band alignment.
\end{abstract}

\maketitle  

Rod-shaped semiconductor nanocrystals~\cite{Alivisatos2000} represent
a class of nanostructures in which the optical and electronic
properties can be tuned by changing the composition, dimensions and
shape, offering an ideal model system to study fundamental properties
and in particular, the transition between $0$ and $1$-dimensional
confinement.  Recent developments in the fabrication of core/shell
seeded
nanorods~\cite{Talapin2003,Manna2007,Talapin2007,Dorfs2008,Sitt2011}
have provided an additional knob by which the electrons/holes can
either be confined to the core region or the shell.  This leads in
some cases to a desired intrinsic charge separation~\cite{Costi2009}
useful for optocatalytic devices.\cite{Costi2008,Amirav2010} In other
cases, where both electrons and holes are confined to the same region,
the nanostructures show remarkable bright and stable
fluorescence.\cite{Manna2007,Talapin2007} These unique features of
semiconductor nanorod heterostructures hold the promise to advance
future light harvesting devices.

Perhaps the most studied of the core/shell nanorod structures is that
of CdSe core with a CdS shell.\cite{Talapin2003} The hole is known to
be localized at the CdSe core due to the large valance band offsets
between CdSe and CdS. On the other hand, the conduction band offsets
are quite small in the bulk, and thus, the electron can either be
localized at the core or at the shell, depending on the size of the
core, leading to a possible transition from a type-I to a
quasi-type-II band alignment.  This has been the focus of numerous
experimental and theoretical studies.

Early work using lifetime measurements combined with model
calculations hinted to a flat band alignment in which case the
electron is confined to the CdS shell and the system is considered to
be a quasi-type-II, regardless of the size of the CdSe
core.\cite{Muller2004,Muller2005} More recent experiments based on
scanning tunneling spectroscopy (STS) analyzed by a simple effective
mass model suggested a conduction band offset of $0.3$eV, which in
principle, would lead to a transition from type-I to quasi-type-II
band alignment as the size of the CdSe core
decreases.\cite{Steiner2008} This apparent controversy has been
addressed using multiexciton spectroscopy (MES) for nanorods with
different core sizes, confirming that a transition from type-I (where
the electron is localized at the core) to a quasi-type-II (where it is
localized at the shell) occurs for a core diameter of
$2.8$nm.\cite{Sitt2009}

Of course, a direct comparison between the STS and MES measurements
must be done with care, as the former ignores the interactions between
the electron and the localized holes, and thus may lead to a more
diffuse electronic state compared to the excitonic state. In fact, the
magnitude of this effect has not been addressed so far and will be
discussed herein.  To add more confusion, this debatable problem has
been revisited very recently using time resolved photoluminescence and
transient absorption spectroscopies,\cite{She2011} showing that the
radiative recombination rate is independent of the CdSe core size,
consistent with a quasi-type-II band alignment for all the system
sizes studied (i.e., cores above $2$nm).  Other recent experimental
studies have also revealed the spatial distribution of the wave
function~\cite{Raino11,Kunneman13} and to long-lived exciton states in
CdSe/CdS dot-in-rod structure.\cite{Wu2013}

\begin{figure*}[t]
\includegraphics[width=0.41\textwidth]{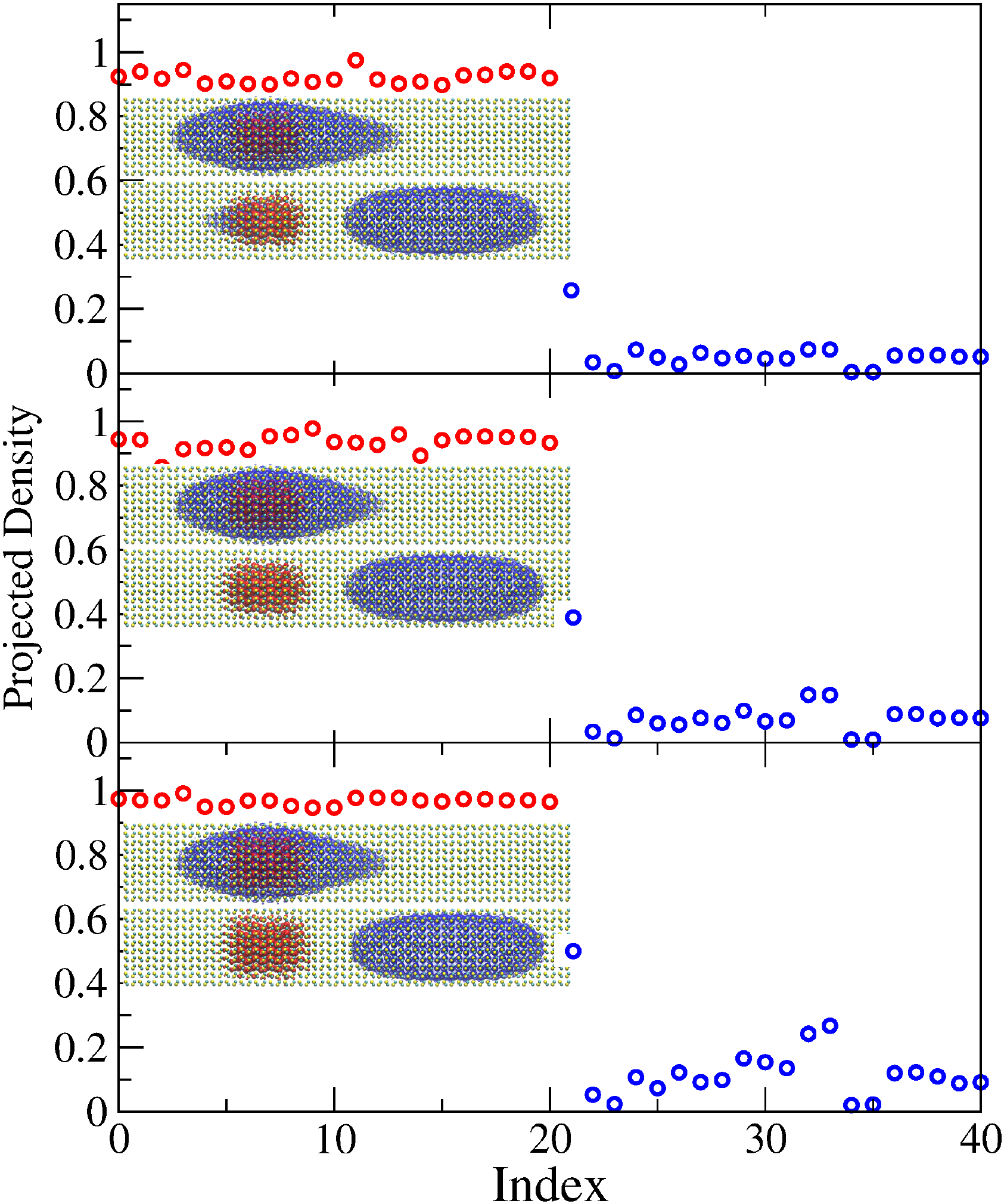}
\includegraphics[width=0.4\textwidth]{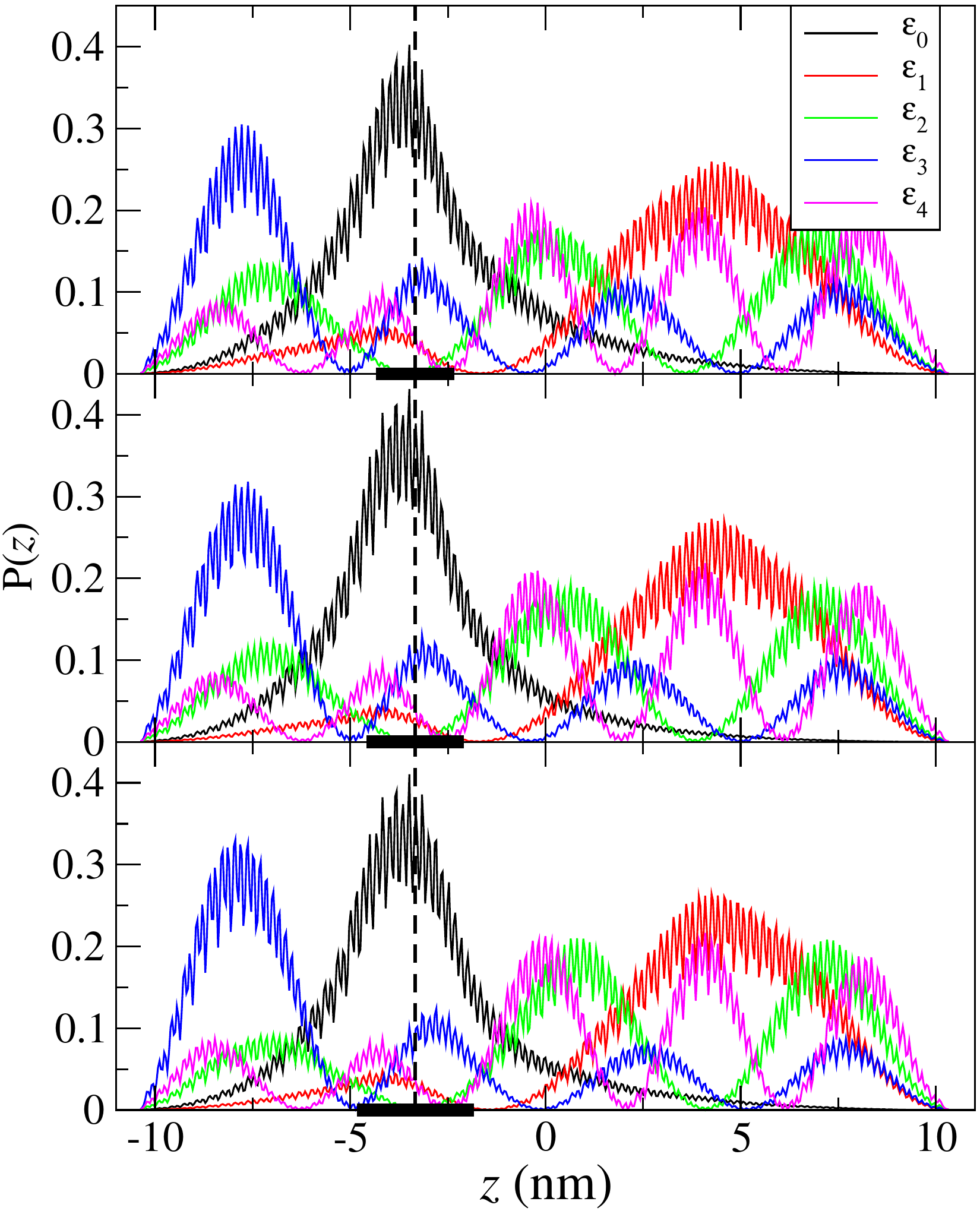}
\caption{Integrated projected hole (red circles) and electron (blue
  circles) densities onto the core (left panels) and the corresponding
  electron density for the $5$ lowest electron level (right) for a $4
  \times 20$~nm CdSe/CdS seeded nanorod.  The seed diameters (from top
  to bottom) are $2$, $2.5$, and $3$~nm.  The dashed vertical line
  shows the position of the center of the core and the solid black
  line superimposed in the x-axis represents core region, which is
  centered at $z=-6.67$~nm.  Insets (left panels) show the hole (red)
  and electron (blue) density isosurfaces for the valance band maximum
  and the two lowest conduction band minimum.}
\label{fig:4nm}
\end{figure*}

This controversy has also attracted numerous theoretical and
computational studies.  Using first-principle calculations, Luo and
Wang~\cite{Luo2010} have examined the band alignment of a CdSe/CdS
core-shell seeded nanorod (dimensions $4.3 \times 15.5$nm) for a core
diameter of $3.4$nm, finding that the hole is localized inside the
CdSe core and the electron in the CdS shell. Since the core size is
larger than the expected transition ($2.8$nm) it contradicts some
measurements and calls for a systematic study of the band alignment
with the core size.

An alternative approach based on an effective mass model was recently
developed by Shabaev {\em et al}.~\cite{Shabaev2012} for "giant"
CdSe/CdS core-shell nanostructures. They find that the Coulomb
potential created by strongly confined holes plays an important role
in the electron confinement, a point overlooked by previous studies.
Shabaev {\em et al}. also examined the effects of core size and
CdSe/CdS conduction band offset dependencies on the electronic
properties of the nanorod.  They found that depending on the band
offset, a transition from type-I to quasi-type-II may occur.  However,
their approach does not provide a quantitative prediction of the actual
band offset in CdSe/CdS nanostructures.

In this work, we calculate the electronic structure of CdSe/CdS
core/shell nanorods of $20$~nm length. We consider two different
values of rod diameter, $4$~nm and $6$~nm, and a number of different
core sizes ($2-4.5$~nm). The rods are faceted, and the spherical CdSe
core was placed at $1/3$ of the length of the rod. The configurations
used for the calculations are equilibrium structures relaxed with
molecular dynamics runs of $100$~ps duration at a temperature of
$300$K. For these runs, interactions between atoms were described by a
recently developed force-field,\cite{Gruenwald2012} which has been
shown to accurately describe CdSe/CdS
heterostructures.\cite{Gruenwald2012,Gruenwald2013} The final
configuration was quenched to remove structural effects of thermal
fluctuations. (See SI for a detailed description of simulation
methods.)

The electronic structure calculations of the seeded nanorods were
performed within the local version of the semiempirical
pseudopotential model,\cite{Zunger94b,Zunger96g,Zunger2000} where the
local screened pseudopotentials were fitted to reproduce the
experimental bulk band structure, band gaps, effective masses, etc.
Furthermore, ligand potentials were used to represent the passivation
layer.\cite{Rabani99} For cadmium and selenium atoms we have used the
pseudopotential developed for CdSe~\cite{Rabani99} while for sulfur
atoms we have fitted the bulk properties of CdS using the existing
pseudopotential for cadmium atoms.\cite{sulfur} The pseudopotential
for CdS gives a flat band alignment with CdSe in the
bulk.\cite{Peng1997} The filter-diagonalization
technique~\cite{Rabani2002a} was then employed to filter nearly $40$
single particle states near the bands edge.  These states were then
used to solve the Bethe-Salpeter equation. In order to compare the
results to the non-interacting case, we used only one hole state, such
that all excitations are associated with electron transitions. We
verified that including more hole states does not affect the electron
density but adds more excitations associated with hole transitions.

In the left panels of Fig.~\ref{fig:4nm} we plot the integrated
projected valance (red circles) and conduction (blue circles)
densities onto the core for $\approx 20$ levels below the valance band
maximum and $\approx 20$ levels above the conduction band minimum.
The results are shown for a fixed shell size ($4 \times 20$~nm) and
for core diameters that vary between $2$~nm and $3$~nm from top to
bottom.  The inset in each panel shows the hole density of the top of
the valance band (red isosurface) and electron density of the two
lowest states of the conduction band (blue isosurface), all
superimposed on the nanorod frame.

\begin{figure*}[t]
\includegraphics[width=0.41\textwidth]{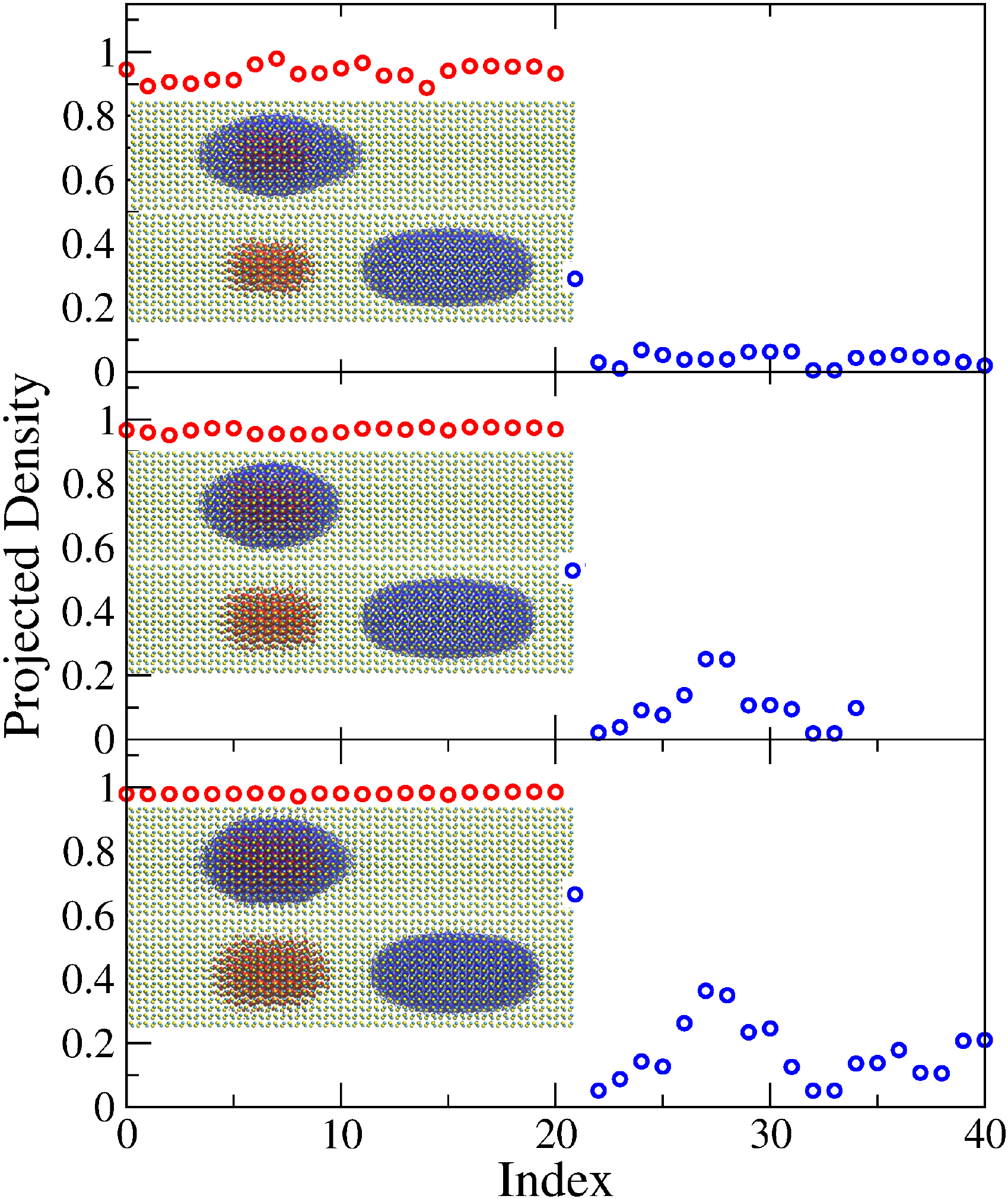}
\includegraphics[width=0.4\textwidth]{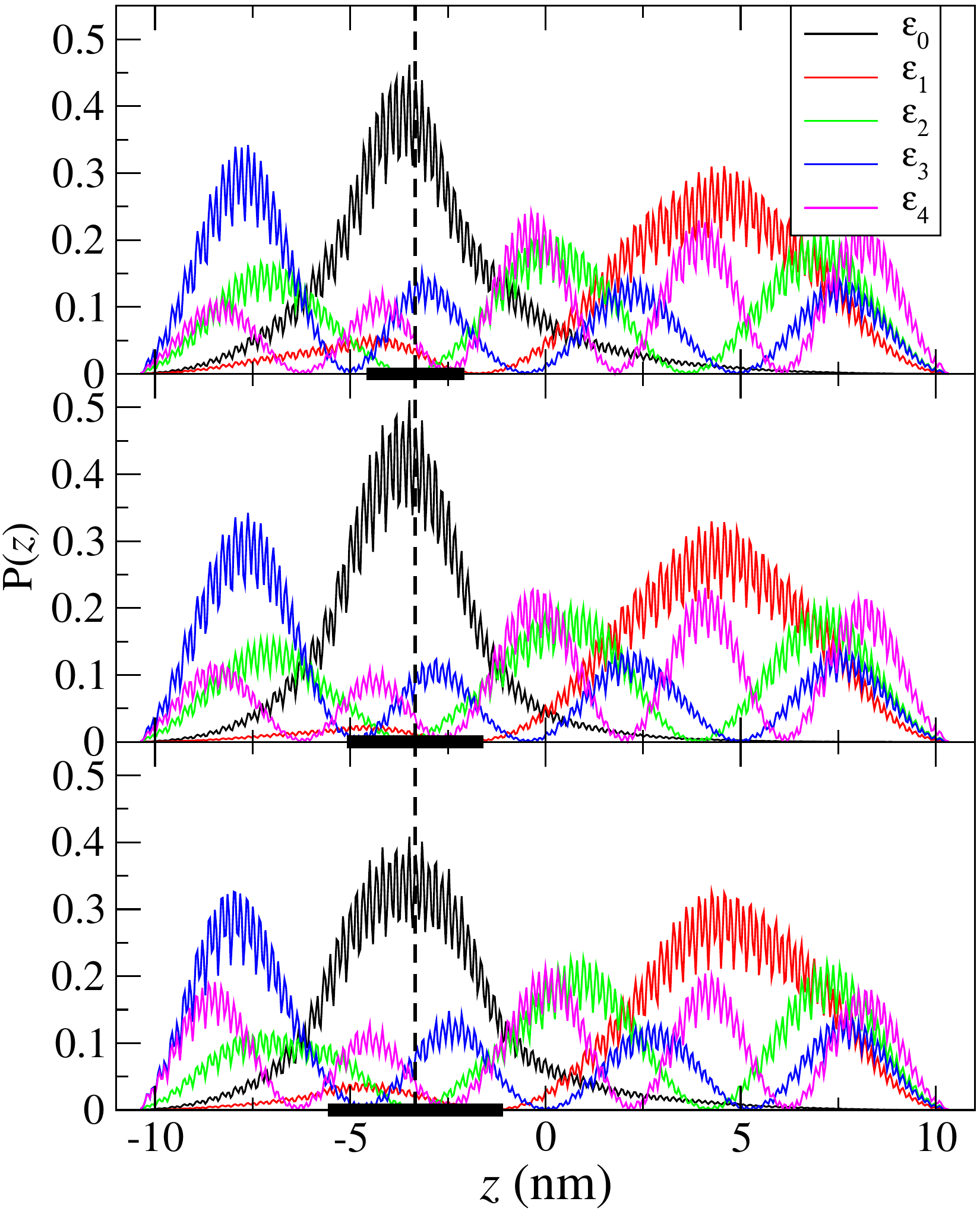}
\caption{Same as Fig.~\ref{fig:4nm} but for a $6 \times 20$~nm
  CdSe/CdS seeded nanorod with seed diameters (from top to bottom) of
  $2.5$, $3.5$, and $4.5$~nm.}
\label{fig:6nm}
\end{figure*}

We find that for all valance states calculated, the value of the
integrated projected density of the hole is close to unity, implying
that the hole is localized in the core with a small probability to
leak into the shell region.  By contrast, the overlap of the lowest
conduction states with the core region depend markedly on core
size. For the smallest core considered here ($2$~nm), we find
appreciable, but minor, core-overlap of electron density only for the
lowest conduction state. All higher states are delocalized in the
shell region. With increasing core size, the core-overlap of the
lowest state, as well as of a number of higher states, increases
markedly, as could be expected from a transition from type-II to
type-I band alignment.

The observed increase in core-overlap, however, is not primarily
caused by a substantial change in band alignment. In the right panels
of Fig.~\ref{fig:4nm}, we plot the electron densities of five lowest
conduction states, projected onto the nanorod axis. As the core size
increases, these densities change only little, as can also be seen in
the insets of the left panels of Fig.~\ref{fig:4nm}. The biggest
contribution to the observed increase in core-overlap thus stems from
the increasing core-volume itself.  This result indicates that the
transition between type-I and type-II behavior is a gradual one. The
threshold value of core size at which different behavior is observed
will therefore likely depend on the nature of the experimental
measurement.

We obtain similar result for a thicker nanorod of same length ($6
\times 20$~nm) with core sizes ranging from $2.5$ to $4.5$~nm (see
Fig.~\ref{fig:6nm}). Like in the case of the thinner nanorod, all
calculated valance states are highly localized in the core region.
Conduction states show an increase in core-overlap with increasing
core size. Comparing rods of different diameter, we find that the
overlap of the lowest conduction state with the core decreases as the
shell diameter increases, consistent with the decrease of the
conduction level with increasing nanorod diameter.

This quasi-particle picture is consistent with recent low-temperature
STS measurements, suggesting that the electron is somewhat delocalized
in the nanorod with a notable amplitude in the core
region.\cite{Steiner2008} However, a direct comparison of our
predictions with results from STS should be carried out with care:
Since the typical broadening in the scanning tip ($\approx 100$~meV)
is larger than the level spacing in the conduction band ($\approx
30$~meV), the transmission is likely to occur through a combination of
conduction states rather than a single state. This makes a direct
comparison of the degree of localization quite hard. Yet, our results
for the lowest conduction state agree well with the experimental
observation that tunneling through the lowest conduction state
diminishes as one moves the tip away from the region of the
core.\cite{Steiner2008}

The single-particle picture discussed in Figs.~\ref{fig:4nm} and
\ref{fig:6nm} might be substantially modified when electron-hole
interactions are taken into account.\cite{Shabaev2012} In fact, most
experimental studies on the band alignment of CdSe/CdS core/shell
nanorods are based on optical measurements in which an exciton is
formed and the magnitude of the electron-hole interactions provides a
measure of the degree of localization.  This is particularly important
for situations when the hole is localized at the core and can bind
strongly the electron, thereby increasing its overlap with the
core.\cite{Shabaev2012} Furthermore, this effect will be strongest for
small cores and might therefore be expected to substantially shift the
value of core size at which the type-I/type-II transition is observed.

\begin{figure*}[t]
\includegraphics[width=0.45\textwidth]{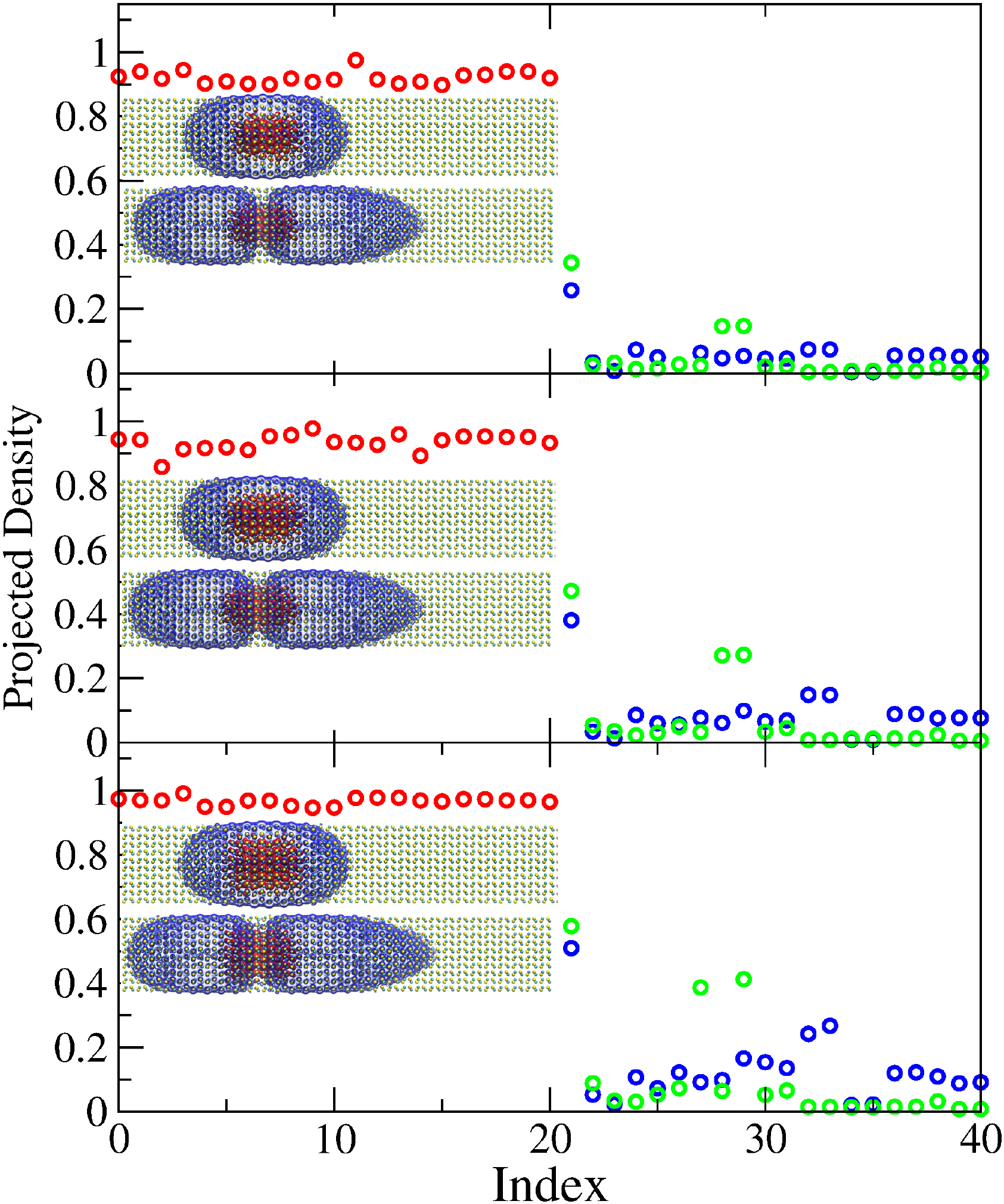}
\includegraphics[width=0.45\textwidth]{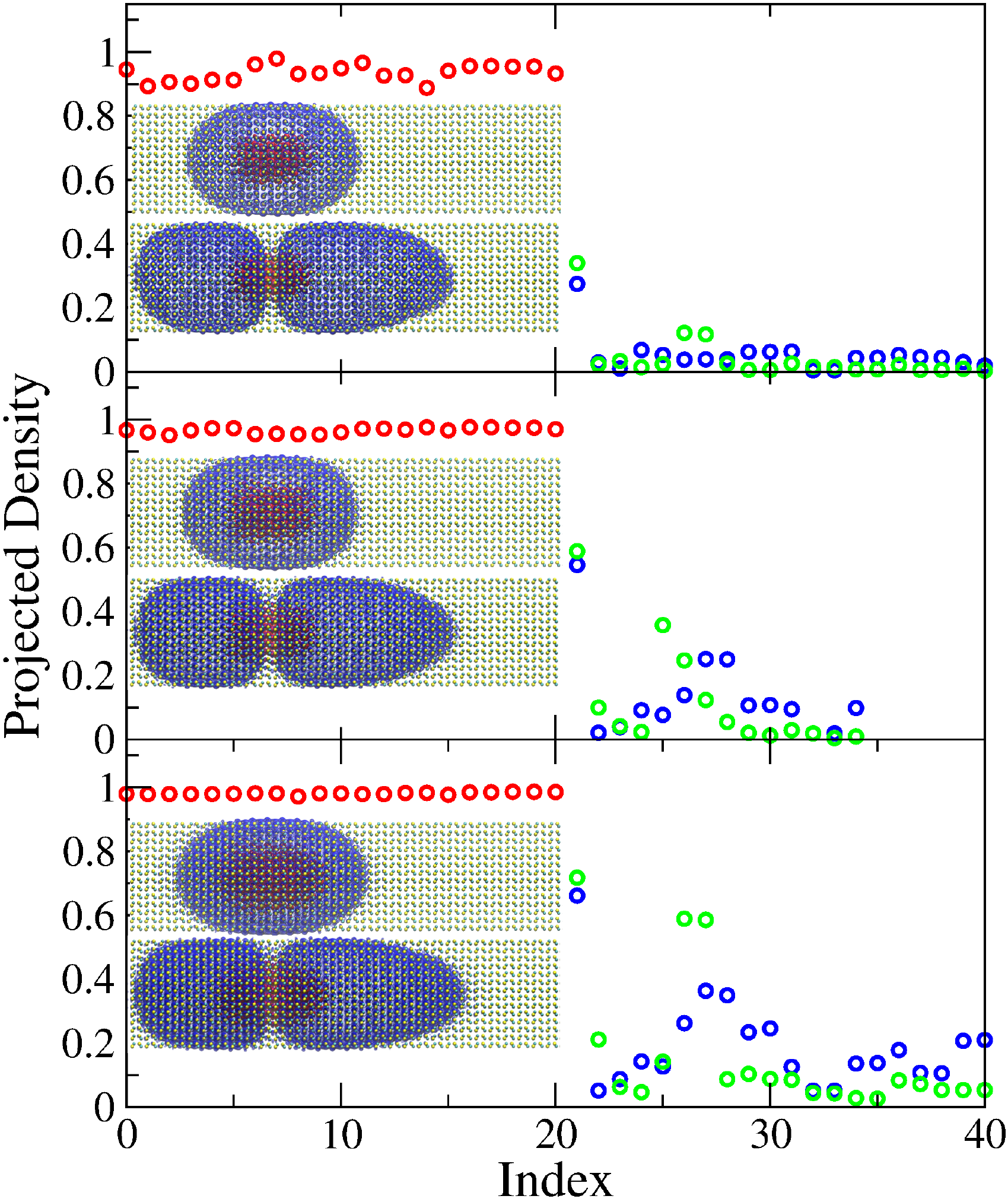}
\caption{Projected electron densities obtained from the BSE for a
  CdSe/CdS seeded nanorod (green circles) along with the projected
  valance (red circles) and conduction (blue circles) densities for
  the noninteracting case.  Left and right panels show results for
  $4\times20$~nm and $6\times20$~nm, respectively.  The core diameter
  (from top to bottom) is $2.5$, $3$, and $3.5$~nm for the left panels
  and $2.5$, $3.5$ and $4.5$~nm for the right panels.  Corresponding
  insets show the hole density (red) and the electron density (blue)
  for the two lowest excitonic state associated.}
\label{fig:bs}
\end{figure*}

Including the interactions between the electron and the hole is a
subtle issue for nanorods, since perturbative techniques that work
well for spherical nanocrystals often fail in nanorods as a result of
small level spacing.\cite{Shabaev2012,Shabaev2004} Here, we resort to
the Bethe-Salpeter approach~\cite{BSE} within the static screening
approximation, where excited states are obtained by diagonalizing the
Bethe-Salpeter equation (BSE) with an exciton Hamiltonian give
by:\cite{Rohlfing2000}
\begin{equation}
\begin{split}  
\nonumber H_{\alpha\equiv{a,i},\beta\equiv{b,j}} &= (\varepsilon_a -
\varepsilon_i) \delta_{ab}\delta_{ij} \\ &- \left(2 \langle \phi_a
\phi_b | W | \phi_i \phi_j \rangle - \langle \phi_a \phi_i | V |
\phi_b \phi_j \rangle\right)
\end{split}
\end{equation}
where $|\phi_{t}\rangle$ are the single particle states with energies
$\varepsilon_t$; $a,b$ label virtual states and $i,j$ occupied states;
$W$ and $V$ are the screened and bare Coulomb potentials,
respectively.

In Fig.~\ref{fig:bs} we show the results for the projected electron
density onto the core obtained from the BSE (green circles) for the $4
\times 20$~nm (left panel) and $6 \times 20$~nm (right panel).  For
comparison, we also include the integrated projected valance (red
circles) and conduction (blue circles) densities onto the core for the
noninteracting case shown in Fig.~\ref{fig:4nm} and
Fig.~\ref{fig:6nm}.

Comparing the maximal value of the electron projected density for the
lowest excitonic state, it is clear that including the electron-hole
interactions leads to an increase of the overlap of the electron wave
function with the core, as one might expect.  A pronounced effect is
also seen for higher excited states. We find two additional excitonic
states that are highly localized near the core region (particularly
for the larger seeds).  These states may well be associated with
recent reports on spatially separated long-lived exciton states in
CdSe/CdS nanorods.\cite{Wu2013}

The observed changes in the degree of core-overlap are partly due to
marked changes in the shape of electronic densities.  The insets for
each panel in Fig.~\ref{fig:bs} show the hole density of the top of
the valance band (red isosurface) and electron density of the two
lowest states of the conduction band (blue isosurface), all
superimposed on the nanorod frame.  Similar to the non-interacting
cases shown in the insets of Figs.~\ref{fig:4nm} and \ref{fig:6nm},
the isosurface of the lowest state is centered around the core region.
The inclusion of electron-hole interactions via the BSE leads to a
somewhat tighter electron density around the seed.  The density of the
second lowest state, however, is profoundly modified from the
non-interacting cases, where the electron is mostly confined to the
shell region away from the core. In the interacting case we observe a
dumbbell-like shape of the isosurface, centered around the core region.

The general trend of increasing core-overlap with core-size remains
unchanged by introducing electron-hole interactions. Like in the
non-interacting case, the main contribution to this increase is the
increasing core-volume itself, rather than a substantial shift in band
alignment.  Furthermore, we do not observe a strong effect of
core-size on the strength of electron-hole interactions, which would
lead to increased localization of electrons in the core primarily for
small cores.

\begin{table*}
\begin{center}
\begin{tabular*}{140mm}{@{\extracolsep{\fill}}c|c|c|c|c|c|c|c|c}
$D_{\rm shell}$ & $D_{\rm core}$ & $E_v$ & $E_c$ & $E_g$ & $E_{ex}$ &
  $\Delta E_{ex}$ & $\tau (\mbox{ns})$ & $f$
  \\\hline
  4 & 0   & -6.60 & -4.01 & 2.59 &      &       &      &  \\
  4 & 2   & -6.38 & -4.10 & 2.28 & 2.14 & 0.46  & 33 & 0.94\\
  4 & 2.5 & -6.31 & -4.10 & 2.21 & 2.05 & 0.35  & 31 & 1.24\\
  4 & 3   & -6.25 & -4.10 & 2.15 & 2.00 & 0.23  & 30 & 1.50\\

  6 & 2.5 & -6.35 & -4.20 & 2.15 & 2.00 & 0.39  & 39 & 0.88\\
  6 & 3.5 & -6.25 & -4.21 & 2.04 & 1.90 & 0.27  & 37 & 1.28\\
  6 & 4.5 & -6.15 & -4.21 & 1.94 & 1.82 & 0.23  & 36 & 1.66\\
  \hline
\end{tabular*}
\end{center}
\caption[]{Calculated energies (in eV) for the top of the valance band
  ($E_v$), bottom of the conduction band ($E_c$), quasi-particle band
  gap ($E_g$), first exciton energies obtained within the BSE
  ($E_{ex}$), and the energy difference between the first exciton in
  the core-only and core/shell structures ($\Delta E_{ex}$) for
  CdSe/CdS core/shell seeded nanorods of different dimensions (in
  nm). $f=\frac{4 m_e E_{0}}{3 \hbar^2 e^2} {\bf \mu}^2$ is the
  oscillator strength for the lowest exciton transition of energy $E_0$
  with transition dipole ${\bf \mu}$ and $\tau$ is the radiative
  lifetime (assuming that for the lowest transition the index of
  refraction is close to $1$~\cite{Rabani2012}).}
\label{ta:ene}
\end{table*}

In Table~\ref{ta:ene} we summarize the relevant energies calculated
for the relaxed nanorods configurations.  The quasi-particle gaps and
the exciton energies for the seeded nanorods are slightly lower
($\approx 0.1$~eV) compared to experimental results.\cite{Sitt2009}
This is likely due to the fact that the pseudopotential used in the
electronic structure calculations was optimized for a perfect lattice
configuration in the neat CdSe and CdS bulk systems, while our
calculations were performed for a relaxed hetero-structure.  The small
overall change in the band gap with the core size (roughly $0.2$eV
going from $D_{\rm core}=2.5$ to $4.5$~nm) is in agreement with the
photoluminescence experiments.\cite{Sitt2009} The shift in the
position of the hole level is larger than that of the electron since
the latter overlaps the shell, which is kept fixed in this process.
This is also consistent with STS measurements.\cite{Steiner2008} We
find that the conduction band minimum is always below that of a neat
CdSe NC (results not shown here)~\cite{Rabani99} for all sizes
studied, confirming a very shallow band offset between the core and
the shell for the electrons.

\begin{figure}
\includegraphics[width=0.4\textwidth]{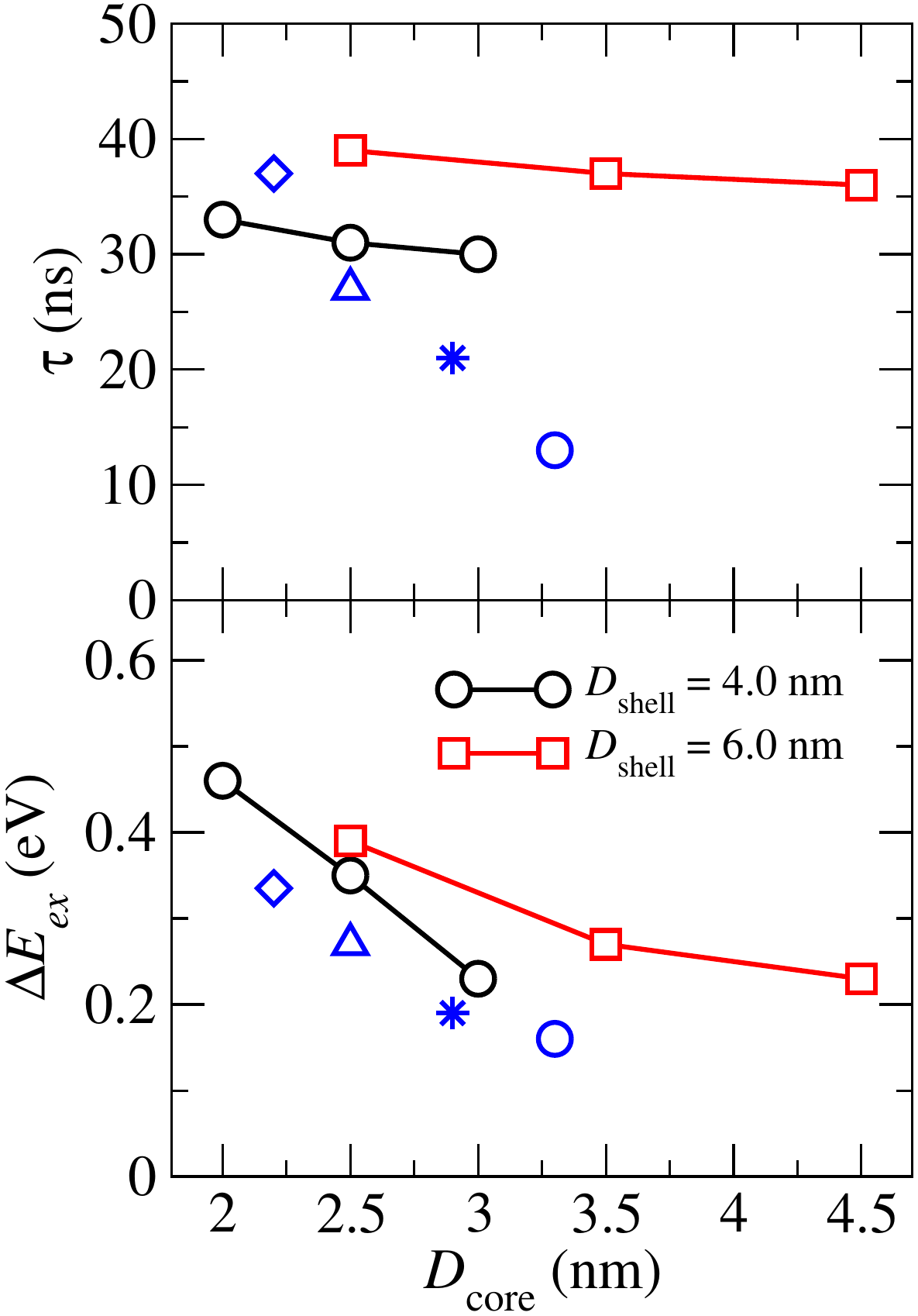}
\caption{Plots of $\Delta E_{ex}$ (lower panel) and $\tau$ (upper
  panel) versus core diameter. Blue symbols are experimental results
  taken from Ref.~\cite{Raino11}. The shell thickness is $4.7 \pm
  0.5$, $4.8 \pm 0.5$, $4.5 \pm 0.5$ and $4.1 \pm 0.4$~nm for the blue
  diamond, triangle, star and circle, respectively.}
\label{fig:energy}
\end{figure}

Table~\ref{ta:ene} also shows the energy difference between the first
exciton in the core only and core/shell structures ($\Delta E_{ex}$)
and the radiative lifetime ($\tau$).  These are also plotted in
Fig.~\ref{fig:energy}.  These quantities depend weakly on the length
of the nanorod, but show a pronounced effect with the core diameter
and rod diameter. The calculated values are compared with the
experimental results for the shift in the absorption peak and the
radiative lifetime,\cite{Raino11} a comparison that can serve to
further assess the accuracy of the model.  The experimental energy
shifts vary from $0.34$~eV to $0.16$~eV when the core size increases
from $2.2$~nm to $3.3$~nm in diameter. The theoretical values are
slightly larger than the corresponding experimental results. However,
our model captures the general trends with core size and the slope of
$\Delta E_{ex}$ is similar. The comparison is complicated by the fact
that experimental values were obtained for nanorods with different
(shell) diameters, and no uncertainties are provided in
Ref.\cite{Raino11} for the values of core diameter (or how these
values were determined).  Assuming a standard deviation of $0.5$~nm,
as quoted for the values of shell diameter, our results agree
quantitatively with experiments.

Comparing the radiative lifetimes to the experimental values, we find
that in both cases $\tau$ decreases with increasing core diameter.
The lifetimes are very similar for the smaller cores (on the order of
$30$~ns), while for the larger cores, we overestimate the lifetimes by
more than $50\%$.  The discrepancy may indicate that for larger cores
there are defects located at the interface between the two
semiconductors which are not included in our model and may affect the
slope of the lifetime with core diameter.  Alternatively, the
experimental lifetimes may include contributions from non-radiative
decay, again not included in the theory. The non-radiative decay is
expected to be more significant for large cores as the density of
phonon changes with the core volume.

Finally, Table~\ref{ta:ene} also provides the values of the oscillator
strength, which increases rapidly with the core size and decreases
with the nanorod diameter.  The dependence on the core and rod
diameters can be explained in terms of the extent of electron
localization and the effect of the core and nanorod dimensions on the
overlap of the electron wave function with the core, as reported
above.

In summary, we have used a combination of molecular dynamics and
electronic structure simulation techniques to study the electronic
properties of CdSe/CdS core/shell seeded nanorods.  For all system
sizes considered here, we find strong localization of the hole in the
core of the nanorod. The overlap of conduction states with the core
region depends on core size: For small cores, minor overlap occurs for
the lowest conduction state only; for larger cores, sizable overlap
is observed for a number of conduction states. When electron-hole
interactions are taken into account, the core-overlap of conduction
states increases but the same trend with core size are observed, which
is consistent with experimental observations of a transition from
type-I to quasi-type-II behavior with decreasing core size. Our
results indicate that this transition is not primarily driven by a
change in band-alignment, but rather by the change in core volume
itself.

{\bf Acknowledgement}
We would like to thank Uri Banin for insightful discussions. HE is
grateful to The Center for Nanoscience and Nanotechnology at Tel Aviv
University for a post-doctoral fellowship. M.G. was supported by the
Austrian Science Fund (FWF) under Grant J 3106-N16.  This work was
supported by the Israel Science Foundation (grant number 611/11).  ER
thanks the Marko and Lucie Chaoul Chair.

\providecommand*\mcitethebibliography{\thebibliography}
\csname @ifundefined\endcsname{endmcitethebibliography}
  {\let\endmcitethebibliography\endthebibliography}{}


\begin{mcitethebibliography}{33}
\bibitem[sul()]{sulfur}
We use the following form for the local pseudopotential in momentum space:
  $V(q)=\frac{a_1(q^2-a_2)}{(a_3 e^{a_4 q^2}+1)}$. The paramters for cadmium
  are $a_1=0.219$, $a_2=0.936$, $a_3=0.196$ and $a_4=1.68$. For selenium:
  $a_1=-0.0291$, $a_2=4.40$, $a_3=-1.20$, and $a_4=0.318$. For sulfur:
  $a_1=-0.0321$, $a_2=4.366$, $a_3=-1.21$, and $a_4=0.282$ in atomic
  units.\relax
\mciteBstWouldAddEndPunctfalse
\mciteSetBstMidEndSepPunct{\mcitedefaultmidpunct}
{}{\mcitedefaultseppunct}\relax
\EndOfBibitem
\end{mcitethebibliography}
\end{document}